# Pairwise comparisons of typological profiles

Søren Wichmann

Max Planck Institute for Evolutionary Anthropology & Leiden University

Eric W. Holman

University of California, Los Angeles

**0. Introduction**[1]

Being rare or 'exotic' is a relative phenomenon. From a Samoan point of view Burushaski is an extremely exotic language, but from the point of view of Telugu much less so. In this brief note we want to look a how different and how similar languages turn out to be in pairwise comparisons and the role that genealogical relatedness plays in this regard. We are interested in knowing whether there is a cut-off point $S_{high}$ in the amount of similarities such that we can be sure that language pairs that have more than $S_{high}$ similarities are all generally thought to be related and also whether there is a cut-of point $S_{low}$ at the other end of the scale such that all languages having less similarities than $S_{low}$ are thought to be unrelated. In other words, if a language is 'normal' relative to some other language (as Burushaski is to Telugu), does this imply that the two languages are related according to commonly accepted classifications? Or, if two languages are mutually very exotic (as Burushaski and Samoan), does this imply that they are not thought to be related in commonly accepted classifications?

The data we use, as well as the genealogical classification, are from the *World Atlas of Language Structures* (Haspelmath et al., ed., henceforth WALS). The conclusions must of course be seen in relation to this particular dataset. Thus, when we observe a certain amount of typological similarity between two languages, this is strictly and only similarity in terms of the kinds of features investigated in WALS. The dataset includes 134 nonredundant features, each of which has from two to nine discrete values. All of these are quite generic typological features. Our conclusions are also limited to the amount of data available. We have required that for any language pair in our sample there should be 45 or more features attested for both members of the pair (a motivation for this precise number follows shortly). This has limited our sample to 320 languages and 29,810 pairs of languages compared. Among these pairs, there are 1,099 which are related according to the classification used in WALS. Henceforth we substitute 'related' for the more cumbersome 'related according to the WALS classification'. We follow this classification because it seeks to meet a consensus view.

---

[1] We would like to thank Bernard Comrie, Cecil Brown, and Dietrich Stauffer for comments on this manuscript.

## 1. Results

Figure 1 presents the overall results of the investigation. As can be seen, the more similar languages get, the greater the probability is that they are related. The figures on which the curve is based are presented in Table 1. Percent similarity was defined as the percentage of attested features for which both languages have the same value. We have binned language pairs in 5% intervals from 10% to 90% similarity. For the plot in Figure 1 the mean percent similarity in each interval was used. Table 1 gives some additional information: it also shows how many language pairs belong in each interval. This is important for the interpretation of the results, as we shall see shortly.

Before giving our interpretation let us explain why we have chosen the criterion that language pairs should have 45 or more features attested for both languages. It turns out that for a criterion of 30 or more features the curve is rather similar but not quite as steep, showing less dependence between the amount of similarity and the probability of finding related pairs. This indicates that the fewer features one operates with, the more prominent is random sampling variability in percent similarity. When operating with a criterion of 60 or more attested features the curve becomes uneven, indicating that the higher criterion passes too few pairs for stable results. This becomes even more pronounced when the criterion is 75 or more features. Obviously, with a more extended database the number of features taken to be criterial could be raised, but 45 is a number that suits the data available in WALS.

Figure 1. The probability of finding related languages

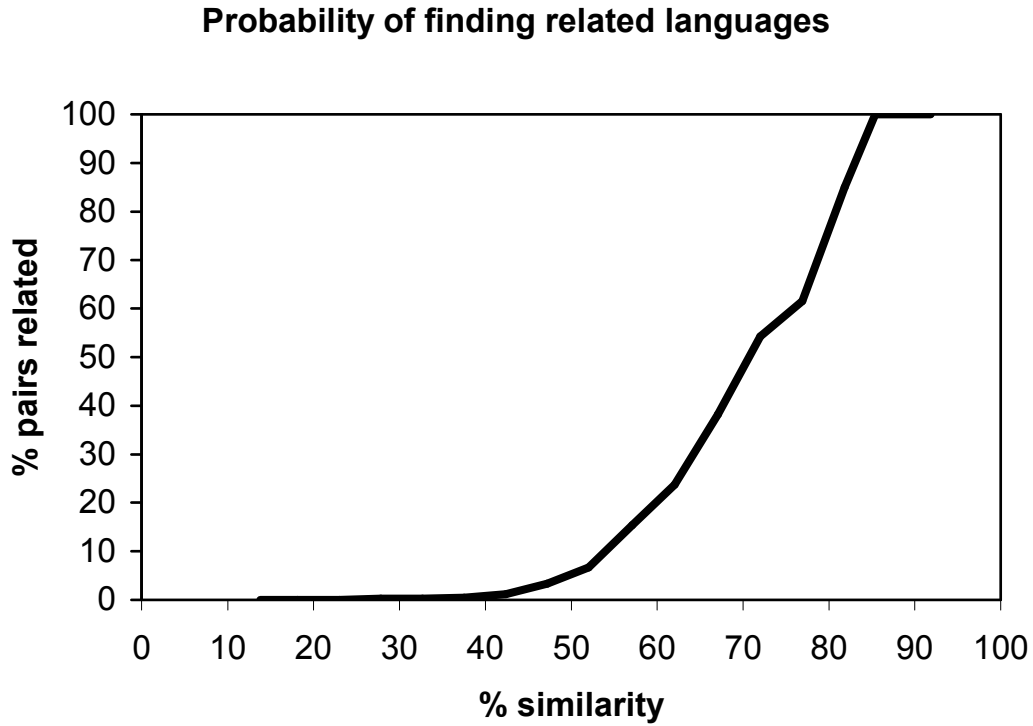

Table 1. Data. (%SIM = % typological similarity between members of pairs; %REL = % of language pairs that are related; PAIRS = number of language pairs in range)

| %SIM | %REL | PAIRS |
|---|---|---|
| 10.0-14.9 | 0 | 11 |
| 15.0-19.9 | 0 | 91 |
| 20.0-24.9 | 0 | 443 |
| 25.0-29.9 | 0.26 | 1566 |
| 30.0-34.9 | 0.33 | 3904 |
| 35.0-39.9 | 0.4 | 6019 |
| 40.0-44.9 | 1.2 | 6772 |
| 45.0-49.9 | 3.26 | 4873 |
| 50.0-54.9 | 6.68 | 3520 |
| 55.0-59.9 | 15.41 | 1551 |
| 60.0-64.9 | 23.72 | 666 |
| 65.0-69.9 | 38.24 | 238 |
| 70.0-74.9 | 54.26 | 94 |
| 75.0-79.9 | 61.54 | 39 |
| 80.0-84.9 | 85 | 20 |
| 85.0-89.9 | 100 | 2 |

| 90.0-94.9 | 100 | 1 |

It may be of interest to mention the language pairs that fall in the lower and upper ranges of the percentage of shared values. Collectors of linguistic trivia may find it interesting that the members of the most divergent language pair in the world (in our dataset), i.e. Tümpisa Shoshone and Wari', are found in the same general area, namely the Americas, that someone who is tired of Romance linguistics should turn to Nivkh and someone fed up with Swedish should visit the Koasatis when looking for something as radically different as it gets. Lists of the 20 most divergent language pairs and the 20 most similar ones are provided in tables 2 and 3.

Table 2. The 20 most divergent language pairs in the sample

| Language A | Language B | Number of features compared | % Similarity |
|---|---|---|---|
| Tümpisa Shoshone | Wari' | 48 | 10.4 |
| Archi | Tukang Besi | 46 | 13 |
| Maybrat | Limbu | 45 | 13.3 |
| Italian | Nivkh | 51 | 13.7 |
| Burushaski | Samoan | 49 | 14.3 |
| Tzutujil | Burmese | 49 | 14.3 |
| Ju\|'hoan | Yup'ik (Central) | 56 | 14.3 |
| Maybrat | Tamil | 55 | 14.5 |
| Nubian (Dongolese) | Acehnese | 48 | 14.6 |
| Swedish | Koasati | 47 | 14.9 |
| Klamath | Wari' | 47 | 14.9 |
| Kongo | Ladakhi | 46 | 15.2 |
| Bashkir | Maori | 46 | 15.2 |
| Berber (Middle Atlas) | Waorani | 45 | 15.6 |
| Lango | Archi | 45 | 15.6 |
| Archi | Thai | 45 | 15.6 |
| Thai | Retuarã | 45 | 15.6 |
| Ijo (Kolokuma) | Kutenai | 50 | 16 |
| Kongo | Evenki | 56 | 16.1 |
| Arabic (Egyptian) | Tümpisa Shoshone | 48 | 16.7 |

Table 3. The 20 most similar language pairs in the sample

| Language A | Language B | Relatedness | Number of features compared | % Similarity |
|---|---|---|---|---|
| Lango | Luo | same genus | 46 | 80.4 |
| Luvale | Zulu | same genus | 97 | 80.4 |
| Khmer | Vietnamese | same family, different genera | 89 | 80.9 |
| Vietnamese | Thai | different families | 110 | 80.9 |
| Khalkha | Tuvan | same family, different genera | 48 | 81.3 |
| Lithuanian | Russian | same family, different genera | 64 | 81.3 |
| Greek (Modern) | Bulgarian | same family, different genera | 64 | 81.3 |
| Khmer | Thai | different families | 91 | 81.3 |
| Polish | Russian | same genus | 71 | 81.7 |
| Russian | Serbian-Croatian | same genus | 45 | 82.2 |
| Swahili | Zulu | same genus | 107 | 82.2 |
| Dagur | Turkish | same family, different genera | 46 | 82.6 |
| Telugu | Kannada | same family, different genera | 47 | 83 |
| Kongo | Nkore-Kiga | same genus | 48 | 83.3 |
| Dutch | German | same genus | 56 | 83.9 |
| Italian | Spanish | same genus | 63 | 84.1 |
| Drehu | Iaai | same genus | 46 | 84.8 |
| English | Swedish | same genus | 60 | 85 |
| French | Italian | same genus | 64 | 85.9 |
| Hindi | Panjabi | same genus | 49 | 91.8 |

While Table 2 does not point in any specific direction and remains a curiosity, Table 3 provides fragments of information which fits into the larger picture that emerges from our study. We note that two pairs of unrelated languages, Vietnamese-Thai and Khmer-Thai, turn up in this list, which otherwise consists of genealogically unrelated language pairs. Furthermore, the rest of the pairs represent a mixture of languages related to different degrees (see Dryer 1992, 2005 for a definition of 'genera').

Returning to Figure 1 and the associated data in Table 1 let us proceed to overall interpretations. We set out asking whether there is some degree of similarity in typological profiles beyond which it is certain that languages are related. The answer is positive, but nevertheless discouraging. Members of language pairs in the sample that are 81.5% or more similar are all related. But only 12 pairs of languages are that similar, in spite of the fact that there are 1099 pairs of related languages in the sample! On the other hand, if there are less than 25% shared feature values all language pairs will be unrelated, and this goes for 545 pairs in the sample. If one allows for a very small margin of error (around 1%), it can predicted that less than 40% shared feature values implies unrelatedness. That goes for 12,034 language pairs in the sample—close to half of the total of 29,810. Thus, lack of similarity is a good predictor of unrelatedness, but presence of similarity is a bad predictor of relatedness.

**2. Are there ways of improving the results?**

We next consider the question of whether the prediction of relatedness could be improved somehow. In other studies (Holman et al. 2006a,b, Brown et al. 2006) we have made exact quantitative explorations of the relationship between typological similarity and geographical distance among languages. Not surprisingly, the greater the geographical proximity is between languages, the more similar they tend to be (this goes for both related and unrelated languages). If one takes into account the areal factor, this might move the cut-off point to allow more accurate predictions of relatedness. Testing this strategy was unsuccessful. We were not able to obtain markedly different results by adjusting the measure of similarity relative to geographical distance: the correlation between adjusted and unadjusted measures was 0.96. The reason for this is probably that the distance measure, as given in the WALS database, identifies the location of a given language (roughly) with its center of extension. This means that some neighbouring languages, such as German and Dutch, are treated as having a certain distance between them when in reality they don't have any. The more widespread the languages compared are, the bigger this problem gets. Since it is impossible to provide adequate measure of geographical distances for 29,810 language pairs, and not just take recourse to a mechanical measure of distance from one WALS dot to another, it is not viable to improve on the cut-off point in such a way.

Also, the 134 features differ appreciably in the distribution of rarity and commonness among their values. It is possible to imagine that taking into account the relative rarity of feature values might improve the predictions. We again failed to obtain markedly different results by adjusting the measure of similarity relative to differences among features: the correlation between adjusted and unadjusted measures was 0.98. The probable reason is that differences among features tend to average out in a sample of at least 45 attested features.

Another strategy to try to improve the power of prediction concerning relatedness would be to weight different features or values of features according to their stability. We have explored ways of measuring stability and have come out with a ranked order of stability for WALS features (Wichmann et al. 2006). Conceivably, if the features shared among languages were weighted for their stability the cut-off point could be pushed a bit. We expect, however, that the results would be similar to the results for taking into account rarity, since stable and unstable features would also average out.

A final strategy to improve the results would be to take into account the areality of features. The linguistic typological literature abounds with statements concerning the susceptibility to diffusion of certain features as opposed to others. In practice, however, it turns out to be virtually impossible to define areas and measure areality in a consistent way. A major contribution of WALS has been to show that most typological features are 'areal' to various extents. Browsing the maps will make it clear to anyone that almost any feature can spread and that whatever features diffuse are the features that happen to exist in an area. Thus, 'areality' is not amenable to quantification in any straightforward way.

**3. Deviant language pairs**

The results reported on in Figure 1 and Table 1 show that there are a few pairs of languages which are related even though showing less than 40% similarities, which is the point where pairs tend overwhelmingly not to be related. It serves the record to provide a list of the pairs of related languages that are deviant in the sense that they show less similarities than related languages normally do. This list is provided in Table 4.

Table 4. Related languages that have unusually different typological profiles (less than 40% similarities)

| Language A | Language B | Language family | Number of features compared | % Similarity |
|---|---|---|---|---|
| Luvale | Ijo (Kolokuma) | Niger-Congo | 52 | 28.8 |
| Zulu | Ijo (Kolokuma) | Niger-Congo | 52 | 28.8 |
| Maidu (Northeast) | Tsimshian (Coast) | Penutian | 48 | 29.2 |
| Ngiti | Koyra Chiini | Nilo-Saharan | 47 | 29.8 |

| Yoruba | Ijo (Kolokuma) | Niger-Congo | 51 | 31.4 |
| --- | --- | --- | --- | --- |
| Mundari | Semelai | Austro-Asiatic | 66 | 31.8 |
| Swahili | Ijo (Kolokuma) | Niger-Congo | 50 | 32 |
| Maung | Yidiny | Australian | 81 | 32.1 |
| Mundari | Khmer | Austro-Asiatic | 78 | 32.1 |
| Koyraboro Senni | Murle | Nilo-Saharan | 65 | 32.3 |
| Koromfe | Ijo (Kolokuma) | Niger-Congo | 49 | 32.7 |
| Beja | Margi | Afro-Asiatic | 45 | 33.3 |
| Sango | Ijo (Kolokuma) | Niger-Congo | 51 | 33.3 |
| Nandi | Koyraboro Senni | Nilo-Saharan | 47 | 34 |
| Nandi | Koyra Chiini | Nilo-Saharan | 52 | 34.6 |
| Marathi | Spanish | Indo-European | 52 | 34.6 |
| Margi | Amharic | Afro-Asiatic | 49 | 34.7 |
| Mundari | Vietnamese | Austro-Asiatic | 88 | 35.2 |
| Garo | Cantonese | Sino-Tibetan | 51 | 35.3 |
| Berber (Middle Atlas) | Kera | Afro-Asiatic | 65 | 35.4 |
| Irish | Marathi | Indo-European | 45 | 35.6 |
| Paamese | Acehnese | Austronesian | 45 | 35.6 |
| Limbu | Mandarin | Sino-Tibetan | 45 | 35.6 |
| Mandarin | Bawm | Sino-Tibetan | 76 | 36.8 |
| Ijo (Kolokuma) | Diola-Fogny | Niger-Congo | 46 | 37 |
| Ngiti | Nubian (Dongolese) | Nilo-Saharan | 54 | 37 |
| Miwok (Southern Sierra) | Tsimshian (Coast) | Penutian | 62 | 37.1 |
| Mundari | Khmu' | Austro-Asiatic | 70 | 37.1 |
| Bagirmi | Nubian (Dongolese) | Nilo-Saharan | 64 | 37.5 |
| Beja | Hausa | Afro-Asiatic | 82 | 37.8 |
| Koromfe | Kisi | Niger-Congo | 45 | 37.8 |
| Yidiny | Tiwi | Australian | 90 | 37.8 |
| Limbu | Meithei | Sino-Tibetan | 45 | 37.8 |
| Kera | Amharic | Afro-Asiatic | 50 | 38 |
| Zulu | Yoruba | Niger-Congo | 104 | 38.5 |
| Beja | Kera | Afro-Asiatic | 57 | 38.6 |
| Ngiyambaa | Maranungku | Australian | 74 | 39.2 |
| Malagasy | Acehnese | Austronesian | 56 | 39.3 |
| Ngiti | Nandi | Nilo-Saharan | 48 | 39.6 |
| Lugbara | Lango | Nilo-Saharan | 53 | 39.6 |
| Fur | Ngiti | Nilo-Saharan | 58 | 39.7 |

Experts in the different families involved will surely have good explanations for these deviant cases. In some cases a pair may in reality not belong to the same family, as in the case of large and not altogether uncontroversial families such as Australian and Nilo-Saharan. In other cases, such as the two pairs featuring Marathi, a wide separation both temporally and geographically and interaction with widely different types of languages may conspire to make a related pair stand out as unusually different. In any case, measuring the amount of typological similarity provides a clue that 'something is going on'—either the classification is potentially wrong or heavy language contact is involved. So the method of comparing typological profiles is potentially useful for someone wishing to probe into the behavior of different languages within a proposed family.

## 4. Conclusions

The results reported on in this note were, in part, unsurprising and, in part, unexpected. Figure 1 showed a close correlation between relatedness and typological similarity. This is what we had expected. But we also expected to find some minimal amount of typological similarity among language pairs which would suffice to predict that two languages are related. It turned out to be the case, however, that the amount of similarity required to make this prediction is so high (81.5%) that only few language pairs qualify. In practice, this means that typological features such as those of WALS are not useful for identifying relatedness among languages when it comes to comparisons of single pairs (when groups of languages are compared the situation may be different, but this issue is beyond the scope of this paper). At the other end of the scale we found that typological *dis*similarity is a good predictor of *un*relatedness: with only a small margin of error one can predict that languages which have 60% or more differences are not related according to the WALS classification. Our finding that a certain amount of typological differences can be used to predict that languages are not commonly believed to be related means that typological differences are a yardstick for gauging the limits of the traditional comparative method.

While it was was not surprising to find a correlation between relatedness and the amount of typological differences among language pairs, this finding may nevertheless steer us in new directions. Presumably there is a correlation between the amount of shared basic vocabulary and relatedness as well. If so, the amount of shared basic vocabulary and the amount of typological similarity should also be correlated, and it may even be possible to start considering whether there is such a thing as a 'typological clock' such that the time of separation of languages of a given family may be inferred from the amount of typological differences within the family. The fact that unrelated languages may be as similar typologically as related ones indicates that for a 'typological clock' to work reasonably well, several pairwise comparison should be made. How, in practice, this kind of methodology could be developed would be an item for future research.